
\input harvmac.tex
\Title{\vbox{\baselineskip12pt\hbox{MRI-PHY/24/95}\hbox{IP-BBSR/95-112}
\hbox{cond-mat/9512173}}}
{\vbox{\centerline{A Calogero-Sutherland Type Model For Branched Polymers}}}

\centerline{ Dileep P. Jatkar}
\smallskip\centerline{\it Mehta Research
Institute of Mathematics \& Mathematical Physics,}
\centerline{\it 10 Kasturba Gandhi Marg, Allahabad 211002, India.}
\centerline{and}
\smallskip\centerline{Avinash Khare}
\smallskip\centerline{\it Institute of Physics, Sachivalaya Marg, Bhubaneswar
751005, India}

\vskip .3in
We show that a Calogero-Sutherland type model with anharmonic interactions
of fourth and sixth orders leads to the matrix model corresponding to the
branched polymers. We also show that by suitably modifying this model one
can also obtain N-particle problems which are connected to matrix models
corresponding to the pure gravity phase as well as corresponding to the
transition point between the soap bubble and the branched polymer phase.
\Date{12/95}

\lref\caloI{F. Calogero, Jour. Math. Phys. {\bf 10} (1969), 2191, 2197.}

\lref\caloII{F. Calogero, Jour. Math. Phys. {\bf 12} (1971) 419.}

\lref\suth{B. Sutherland, Jour. Math. Phys. {\bf 12} (1971) 246.}

\lref\ddsw{S. Das, A. Dhar, A. Sengupta and S. Wadia, Mod. Phys. Lett.
{\bf A5} (1990) 1041.}

\lref\bipz{E. Brezin, C. Itzykson, G. Parisi and J. B. Zuber, Commun.
Math. Phys. {\bf 59} (1980) 35.}

\lref\sla{B.D. Simons, P.A. Lee and B.L. Altshuler, Phys. Rev. Lett.
{\bf 72} (1994) 64.}

In recent years, Calogero - Sutherland (CS) type N-body
problems in one dimension have received considerable attention in the
literature \refs{\caloI, \caloII, \suth}. In particular, remarkable
connections have been found between such models and seemingly totally
different models \refs{\sla}. Further, these models are shown to correspond
to ideal gas in one dimension with fractional exclusion statistics. The
purpose of this note is to point out one more such connection.
In particular, we show that the N-body problem with equal mass in
1-dimension characterized by $(\hbar = 2m = 1,\ g > - 1/2, B > 0)$
\eqn\hami{H = - \sum^N_{i=1} {\partial^2\over\partial x_i^2} +\sum^N_{i<j}
{g\over (x_i-x_j)^2} + B\sum_{i} x_i^2 + A(\sum_{i} x_i^2)^2
+ C(\sum_i x_i^2)^3}
is related to the matrix model corresponding to the problem of branched
polymers. We also show that by suitably
modifying this model one can also obtain N-body problems which are connected
to matrix models corresponding to pure gravity phase as well to the transition
point between the pure gravity and the branched polymer phase.

Consider a many-body system with the hamiltonian given in eq.\hami .
The corresponding Schr\" odinger equation is given by
\eqn\sch{H\psi=E\psi}
Following Sutherland\refs{\suth} we will write the wave function as a product
of two wave functions, one of which carries the Jastraw factor i.e. the
antisymmetric part of the wave function and the remaining part which carries
the exponential damping terms
\eqn\wvfn{\psi=\phi(x_i)\varphi(x_i)}
where
\eqn\jastraw{\phi=\prod_{i<j}\mid x_i-x_j\mid^{\lambda}}
and
\eqn\damp{\varphi=\exp{(-\alpha\sum_{i=1}^{N}x_i^2-\beta(
\sum_{i=1}^Nx_i^2)^2)}.}
On substituting this wave function in eq.\sch\ we find the relation between the
power of the Jastraw factor and the strength of the inverse square potential
\eqn\relat{\lambda^2-\lambda=g/2.}
Solving this for $\lambda$ we get
\eqn\vall{\lambda={1\over 2}[1+(1+2g)^{1\over 2}].}
Further, we find that the coefficients in the wavefunction and the coupling
constants appearing in the hamiltonian are related by
\eqn\alp{A=16\alpha\beta,\qquad B=4[\alpha^2 -\beta(N+2+\lambda N(N-1))],
\qquad C=16\beta^2.}
In case these relations are satisfied then the wave function as given by
eq. \wvfn\ is the ground state eigenfunction and the corresponding eigenvalue
is given by
\eqn\eigen{E_0={A\over {2(C)^{1\over 2}}}(N + \lambda N(N-1))}

The ground state eigenfunction of this many body system can be interpreted in
terms of
a matrix model\refs{\suth}. In particular, square of the modulus of this
eigenfunction can be interpreted as the weight function of the corresponding
matrix integral. For our case it reduces to the following matrix integral
\eqn\mat{Z_N=\int dM \exp{(-{1\over 2}{\rm Tr} M^2-{b'\over N}({\rm Tr}
M^2)^2)}}
involving hermitian matrices provided $\lambda=1$. Of course, $\lambda=1$
means, in the original model there in no inverse square interaction.
This in turn makes the many body problem even simpler.

This matrix integral has been studied in detail \refs{\ddsw}. The leading
order solution is given by taking large $N$ limit where the eigenvalues scale
as
\eqn\larn{x_i = \sqrt{N}x(i/\sqrt{N}) = \sqrt{N}x(z)}
with the continuous variable $z=i/\sqrt{N}$ taking values between $0$ and $1$.
In the large $N$ limit we can solve the integral in the saddle point
approximation. The density of eigenvalues $u(x)$ is defined as
\eqn\dens{u(x)={dz\over dx},}
and its second moment is
\eqn\sec{c=\int_{-\mu}^{\mu}dxu(x)x^2.}
Using these definitions the saddle point equation can be written as
\eqn\saddle{{1\over 2}x+2b'cx=P\int_{-\mu}^{\mu}dy{u(y)\over x-y}.}
Where $P$ stands for principal value.
The eigenvalue range $(-\mu, \mu)$ is determined by normalization of the
eigenvalue density. The solution to the saddle point equation satisfying
the proper asymptotics\refs{\bipz} is
\eqn\sdl{u(x)={1\over\pi}({1\over 2}+2b'c)\sqrt{\mu^2-x^2}.}
 From the normalization as well as the self-consistency conditions
\sec\ it then follows that
\eqn\self{b'\mu^4 = 4 - \mu^2.}
The free energy of the model in the large $N$ limit now takes the form
\eqn\free{\eqalign{E(b') &= \lim_{N\rightarrow\infty}\ln Z_N\cr
&=\int dx {1\over 2}u(x)x^2 + b'c^2 - \int dxdy u(x)u(y)\ln\mid x-y\mid }}
from where we obtain
\eqn\ener{E_{0}(b')-E_{0}(0) = {1\over 16}(\mu^2 -4) -{1\over 2}
\log {\mu^2 /4}}.

It is well known that the matrix models have interpretation in terms of
summing over random surfaces. The interpretation in our case is that
the quartic coupling $\beta$ corresponds to the touching random surfaces.
This phase, where there are several
random surfaces touching each other is interpreted geometrically as the
branched polymer phase. It is well known that in the branched polymer phase
of the random surfaces the susceptibility exponent is positive and is given by
\eqn\susc{\gamma_s = 1/2.}

We will not reproduce all the details here but it suffices to say that the
analysis of \refs{\ddsw} can be carried through in a straightforward manner.
Following their analysis \refs{\ddsw} it is easy to see that in our case the
susceptibility exponent
is indeed $1/2$ at $b' =- 1/16$ thereby establishing the connection of our
N-body problem with the matrix model for the branched polymer phase.

If we modify the many body potential by adding the term
\eqn\newpt{V_{new}=A'\sum_{i=1}^{N} x_{i}^4+B'\sum_{i=1}^{N}
x_{i}^6+C'\sum_{i=1}^Nx_i^4\sum_{i=1}^Nx_i^2 +D'\sum_{i<j}(x_i-x_j)^2,}
and suitably adjust the couplings $A'$, $B'$, $C'$ and $D'$ we can
get the susceptibility exponent $\gamma_s =1/3$ as well as $\gamma_s= -1/2$.
These cases correspond to the crossover from soap bubble phase to the branched
polymer phase and the pure gravity phase respectively. For the crossover
phase, the ground state eigenfunction is again as given by eq. \wvfn\ where
$\phi$ is as given by eq. \jastraw\ while $\varphi$ is given by
\eqn\cross{\varphi=\exp(-\alpha\sum_{i=1}^Nx_i^2-\beta(\sum_{i=1}^Nx_i^2)^2
-\gamma\sum_{i=1}^Nx_i^4)}
where $\alpha$, $\beta$ and $\gamma$ are related to the coefficients in the
many body potential as follows
\eqn\rel{\eqalign{B &= 4[\alpha^2-3\gamma-6\lambda\gamma(N-1)-\beta(2+N+
\lambda N(N-1))]\cr
A &= 16\alpha\beta,\quad C=16\beta^2,\quad A'=16\alpha\gamma\cr
B' &= 16\gamma^2,\quad C' = 32\beta\gamma,\quad D' = 4\lambda\gamma.}}
In this case the ground state energy is as given by eq. \eigen .

In the pure gravity case, the ground state eigenfunction is again given by
eq. \wvfn\ with $\phi$ being given as before by eq. \jastraw\ while $\varphi$
is now of the form
\eqn\grav{\varphi=\exp(-\alpha\sum_{i=1}^Nx_i^2-\gamma\sum_{i=1}^Nx_i^4)}
where the relation between $\alpha$, $\gamma$ and the many body potential is
\eqn\grel{\eqalign{A&=C=C'=0,\quad B=4\alpha^2-12\gamma(1-2\lambda(N-1),\cr
A' &= 16\alpha\gamma,\quad B' = 16\gamma,\quad D' = 4\lambda\gamma.}}
The ground state energy is now given by
\eqn\gro{E_0 = {A'\over 2(C)^{1\over 2}}(N+N(N-1)\lambda)}

We thus see that the N-body quantum mechanical problem on a line with the
hamiltonian \hami\ corresponds to the problem of branched
polymers. By adding suitable extra potential \newpt , we find that the new
many body system corresponds to the pure gravity phase as well
as the
crossover of the soap bubble phase to the branched polymer phase.
It would be interesting to see if this relation could be further explored
to get better understanding of the branched polymer phase.
\vfill
\eject
\listrefs
\bye